\documentclass[12pt,letterpaper]{article}

\usepackage[margin=1in]{geometry}
\usepackage{amsmath,amssymb}
\usepackage{array}
\usepackage{booktabs}
\usepackage{multirow}
\usepackage[round,authoryear]{natbib}
\usepackage{indentfirst}
\usepackage{placeins}
\usepackage{setspace}
\usepackage{hyperref}
\usepackage{microtype}
\usepackage{parskip}
\usepackage{graphicx}
\usepackage{float}
\usepackage{longtable}
\usepackage{tabularx}

\hypersetup{
    colorlinks=true,
    linkcolor=blue,
    citecolor=blue,
    urlcolor=blue,
    pdftitle={Bankruptcy Prediction from 10-K Narratives: Evidence from Interpretable Text Scores and Accounting Baselines},
    pdfauthor={Zhen Zhang, Moxuan Zheng, Tongchen Zhang, Luyun Lin, Yiqing Wang, Lixing Lin}
}

\newcolumntype{Y}{>{\raggedright\arraybackslash}X}
\begin{document}

\begin{titlepage}
\centering
\vspace*{1.5cm}

{\Large\bfseries Bankruptcy Prediction from 10-K Narratives: Evidence from Interpretable Text Scores and Accounting Baselines}

\vspace{1.5cm}

\begin{center}

{\normalsize
\begin{minipage}[t]{0.30\textwidth}
\centering
Zhen Zhang\textsuperscript{\normalfont *}\\[4pt]
\textit{Independent Researcher}\\[4pt]
Jersey City, NJ, USA
\end{minipage}
\hfill
\begin{minipage}[t]{0.30\textwidth}
\centering
Moxuan Zheng\textsuperscript{\normalfont *}\\[4pt]
\textit{Independent Researcher}\\[4pt]
Jersey City, NJ, USA
\end{minipage}
\hfill
\begin{minipage}[t]{0.30\textwidth}
\centering
Tongchen Zhang\\[4pt]
\textit{Independent Researcher}\\[4pt]
Jersey City, NJ, USA
\end{minipage}
}

\vspace{0.5cm}

{\normalsize
\begin{minipage}[t]{0.30\textwidth}
\centering
Luyun Lin\\[4pt]
\textit{Citi Group}\\[4pt]
Dallas, TX, USA
\end{minipage}
\hfill
\begin{minipage}[t]{0.30\textwidth}
\centering
Yiqing Wang\\[4pt]
\textit{Citi Group}\\[4pt]
Dallas, TX, USA
\end{minipage}
\hfill
\begin{minipage}[t]{0.30\textwidth}
\centering
Lixing Lin\\[4pt]
\textit{Yale University}\\[4pt]
New Haven, CT, USA
\end{minipage}
}

\end{center}

\begingroup
\renewcommand{\thefootnote}{\fnsymbol{footnote}}
\footnotetext[1]{Corresponding authors: Zhen Zhang (\texttt{billzhangzhen98@gmail.com}) and Moxuan Zheng (\texttt{pollyzheng9527@gmail.com}).}
\endgroup

\vspace{1cm}

\begin{abstract}
Bankruptcy is a low-frequency but high-impact corporate event, making early risk identification important for creditors, investors, regulators, and risk managers. Traditional bankruptcy-prediction models rely primarily on accounting ratios, but these measures may reflect financial deterioration only after it appears in reported financial statements. Narrative disclosures in annual 10-K filings may therefore provide incremental warning signals about emerging distress. This study examines whether 10-K narratives improve bankruptcy prediction beyond conventional accounting variables. Using firm-year observations matched to 10-K text, SEC financial statement data, and bankruptcy events from the Florida--UCLA--LoPucki Bankruptcy Research Database, the analysis evaluates bankruptcy risk over the year following the 10-K filing date. The paper develops a transparent Pre-Bankruptcy Stress (PB Stress) Score, a dictionary-based measure designed to capture distress-specific language related to liquidity and funding stress, debt covenant and refinancing stress, operating deterioration, restructuring and legal distress, and business fragility. The score is evaluated against a five-variable accounting baseline and a Loughran--McDonald dictionary benchmark. In the primary one-year holdout test, adding the PB Stress Score increases AUC from 0.8323 to 0.9019 and raises top-decile bankruptcy capture from 44.12\% to 64.71\%. The positive incremental pattern remains visible across bootstrap inference, alternative accounting benchmarks, alternative outcome definitions, and out-of-time validation. The findings indicate that distress-specific 10-K narratives provide interpretable incremental information for bankruptcy-risk monitoring beyond conventional accounting ratios.
\end{abstract}

\vspace{0.5cm}

\noindent
\begin{minipage}{0.95\textwidth}
\textbf{Keywords:} bankruptcy forecasting; financial distress; 10-K filings; textual analysis; Loughran-McDonald dictionary; interpretable machine learning.

\end{minipage}

\vspace{0.5cm}

\end{titlepage}

\section{Introduction}
\label{sec:intro}

Predicting corporate bankruptcy is important for creditors, investors, regulators, and risk managers because early identification of financial distress supports credit monitoring, portfolio risk management, supervisory oversight, and capital allocation. Bankruptcy can impose substantial losses on lenders and investors, disrupt business relationships, and transmit stress through broader financial and commercial networks. For this reason, a large bankruptcy-prediction literature studies whether observable firm characteristics can distinguish financially vulnerable firms from healthier firms before default occurs. Classic models such as \citet{beaver1966}, \citet{altman1968}, \citet{ohlson1980}, and \citet{zmijewski1984} show that corporate failure risk is strongly related to information contained in firms' financial statements.

Traditional bankruptcy-prediction models rely primarily on accounting ratios, which summarize a firm's financial condition using quantitative relationships derived from financial statement items. These measures typically capture leverage, liquidity, profitability, firm size, and cash-flow capacity. For example, leverage ratios reflect the extent to which a firm depends on debt financing, liquidity ratios assess whether short-term assets are sufficient to cover short-term obligations, profitability ratios capture earnings-generating ability, and cash-flow ratios measure whether operating cash flows can support repayment capacity. Such ratios remain useful because they provide standardized and economically interpretable signals of financial weakness. In general, firms with higher leverage, weaker liquidity, lower profitability, smaller scale, and weaker operating cash flow are more likely to experience financial distress.

However, accounting ratios may not capture all relevant warning signals early enough. Because ratios are constructed from aggregated financial statement numbers, they may update only after deterioration has already appeared in reported accounting outcomes. Narrative disclosures, by contrast, can describe emerging risks, management concerns, and forward-looking uncertainties before these conditions are fully reflected in reported financial ratios. Examples include refinancing pressure, covenant problems, liquidity warnings, going-concern uncertainty, restructuring activity, litigation contingencies, and dependence on major customers or suppliers. These disclosures may therefore provide economically meaningful information about the mechanisms through which financial stress intensifies, even when conventional accounting measures have not yet deteriorated sufficiently to dominate a prediction model.

Form 10-K filings are a natural source of this additional information. A Form 10-K is the annual report filed by U.S.\ public companies with the Securities and Exchange Commission (SEC), and it combines audited financial statements with narrative discussion of the firm's business, risks, operations, liquidity, and capital resources \citep{sec2024}. This study defines the bankruptcy-prediction window from the 10-K filing date because that is the point at which the full annual disclosure becomes publicly available to investors, creditors, regulators, and other outside users.

This study focuses on text extracted from Item 1, Item 1A, and Item 7 of the 10-K. Item 1 describes the business, Item 1A discusses risk factors, and Item 7 contains Management's Discussion and Analysis of financial condition and results of operations. These sections are especially relevant for bankruptcy prediction because they are among the most likely parts of the filing to contain narrative signals about operating weakness, liquidity and capital-resource pressures, debt and covenant problems, refinancing needs, restructuring activity, litigation risks, and going-concern uncertainty.

Prior text-based approaches show that narrative disclosure contains predictive information, but many existing measures are not designed specifically for bankruptcy-related stress. For example, the Loughran--McDonald dictionary captures broad negative, uncertain, or litigious tone in financial text \citep{lm2011}, but it is not designed to isolate distress mechanisms that are especially relevant to bankruptcy prediction. A broad tone measure may therefore detect general negativity without distinguishing whether the language reflects financing pressure, covenant strain, operating deterioration, or other pre-bankruptcy conditions. This limitation motivates the need for a bankruptcy-specific textual measure focused on language directly related to financial-distress mechanisms.

This paper develops a transparent \textit{Pre-Bankruptcy Stress Score} (PB Stress Score) designed to capture bankruptcy-specific narrative warning signals in 10-K filings. The score is constructed from five pillars: (1) liquidity and funding stress, (2) debt, covenant, and refinancing stress, (3) operating deterioration, (4) restructuring and legal distress, and (5) business fragility. The paper then tests whether this distress-specific text measure improves bankruptcy prediction beyond a parsimonious five-variable accounting baseline and beyond a broad text benchmark based on the Loughran--McDonald dictionary. This emphasis on transparency is also consistent with broader use of interpretable machine learning in financial decision making, including portfolio optimization and credit default analysis \citep{song2023,yang2025}.

Empirically, the PB Stress Score adds substantial predictive value. In the primary one-year holdout test, augmenting the financial baseline with the PB Stress Score increases the area under the ROC curve (AUC) from 0.8323 to 0.9019 and raises top-decile bankruptcy capture from 44.12\% to 64.71\%. The positive incremental pattern also remains visible across multiple robustness checks. These findings suggest that structured distress language in 10-K narrative disclosures contains incremental bankruptcy information beyond conventional accounting ratios alone.

This paper contributes to the bankruptcy-prediction and financial-text literature in three ways. First, it aligns the disclosure date and prediction horizon by defining bankruptcy outcomes from the 10-K filing date, which is the economically relevant point at which the annual disclosure becomes publicly observable. Second, it develops a transparent and economically interpretable distress-specific text measure rather than relying only on broad tone indicators. Third, it evaluates the incremental value of this score relative to both accounting baselines and an established financial dictionary benchmark using holdout testing and multiple robustness checks.

\section{Related Literature}
\label{sec:lit}

\subsection{Accounting-Based Bankruptcy Prediction}

The bankruptcy-prediction literature begins with the premise that accounting information contains forward-looking signals of financial distress. \citet{beaver1966} showed that selected financial ratios can distinguish failed and non-failed firms before bankruptcy, and \citet{altman1968} showed that combining ratios improves classification relative to single-variable analysis. \citet{ohlson1980} and \citet{zmijewski1984} remain especially relevant because they provide transparent probabilistic accounting-based benchmarks built around size, leverage, liquidity, profitability, and related balance-sheet conditions.

Later research clarifies why these models should be treated as forecasting frameworks rather than timeless coefficient formulas. \citet{begley1996} and \citet{grice2003} show that classical bankruptcy models may require re-estimation or reinterpretation in later samples. \citet{shumway2001} argues that bankruptcy prediction should be treated explicitly as a forward-looking prediction problem, while \citet{beaver2005} and \citet{campbell2008} show that parsimonious accounting variables remain informative over time.

This literature motivates the financial baselines used in this paper. The Ohlson-style and Zmijewski-style scores provide named benchmark specifications. The primary five-factor accounting model is not presented as a new canonical bankruptcy formula. Rather, it is a transparent accounting-only benchmark built from recurring dimensions in the literature: firm size, leverage, liquidity, profitability, and operating cash flow. Operating cash flow is included because the paper studies one-year post-filing bankruptcy risk, where near-term internal cash generation and funding pressure are economically important \citep{gombola1987}.

\subsection{Supervisory and Practitioner Credit-Risk Frameworks}

The five-factor financial baseline is also consistent with supervisory and practitioner credit-risk frameworks. The Office of the Comptroller of the Currency (OCC)'s \emph{Rating Credit Risk} handbook emphasizes that the primary consideration in credit assessment is the strength of the primary repayment source and calls for rigorous analysis of revenues, margins, income and cash flow, leverage, liquidity, and capitalization \citep{occ2001}. The Federal Reserve's interagency leveraged lending guidance similarly emphasizes sustainable capital structure, reasonable cash flow and balance-sheet leverage, the borrower's capacity to repay and de-lever, covenant protections, liquidity analysis, and the ability to meet debt maturities \citep{fed2013}. More recently, OCC guidance on refinance risk highlights high leverage, constrained liquidity, near-term maturities, and poor financial performance as key drivers of elevated commercial credit risk \citep{occ2024}.

These supervisory themes map directly to the five-factor accounting baseline used here. Firm size proxies for scale and resilience; leverage captures balance-sheet burden; liquidity captures short-term financial flexibility; profitability captures earnings capacity; and operating cash flow captures internally generated repayment ability. The baseline is therefore intentionally parsimonious, but it is not arbitrary or weak. It is anchored in both the bankruptcy-prediction literature and in supervisory credit-risk practice, making it a credible accounting-only benchmark against which to test whether 10-K narrative disclosures add incremental bankruptcy-risk information.

\subsection{Financial Text and Narrative Disclosure}

A second stream of literature examines whether narrative disclosures contain information about firm outcomes and risk. \citet{li2008} shows that annual report readability is related to performance and earnings persistence. \citet{bodnaruk2015} use 10-K text to gauge financial constraints, and \citet{mayew2015} show that MD\&A disclosure is informative about a firm's ability to continue as a going concern. Research on risk-factor disclosures also supports the relevance of 10-K narratives: \citet{kravet2013} link changes in textual risk disclosure to investor risk perceptions, and \citet{campbell2014} show that mandatory risk-factor disclosure contains information about firm risk exposures.

Within financial-text research, dictionary-based approaches remain especially important because they are transparent and replicable. \citet{lm2011} show that general-purpose sentiment dictionaries perform poorly in financial documents, and \citet{lm2016} survey the resulting finance-text literature. Their negative and uncertainty categories therefore provide a natural benchmark in this paper. At the same time, the LM dictionary is a broad financial-text measure rather than a bankruptcy-specific one. Negative and uncertain language may reflect litigation, competition, regulation, macroeconomic exposure, or disclosure style without necessarily indicating near-term bankruptcy risk.

Recent work also applies more flexible textual models to bankruptcy prediction. \citet{mai2019} use deep learning models with textual disclosures and show that text can improve bankruptcy prediction, but their approach emphasizes learned representations and predictive performance. More recent financial NLP work studies retrieval-augmented systems for long financial reports, LLM-based extraction frameworks for structured 10-K disclosures, benchmark-based evaluation of LLM performance on financial quantitative tasks, and multi-LLM architectures for financial sentiment forecasting, illustrating current interest in retrieval, extraction, model evaluation, and semantic signal construction in financial text analysis \citep{cheng2026a,liu2026,kang2026,zhang2026}. The present study instead emphasizes transparency, interpretability, and incremental value relative to accounting baselines and a standard finance-specific dictionary benchmark.

\subsection{Distress-Specific Text Scoring and Interpretability}

This distinction motivates the paper's distress-specific text design. The research question is not whether financial text matters in general, but whether 10-K narrative disclosure about borrower stress mechanisms adds incremental value beyond accounting variables and a standard finance-text dictionary benchmark.

The PB Stress Score is designed for that purpose. It focuses on categories that are conceptually close to near-term bankruptcy risk: liquidity and funding pressure, debt covenant and refinancing stress, operating deterioration, restructuring or workout pressure, and business fragility. These categories are not intended as ad hoc keywords. Rather, they are grounded in the same credit-risk themes emphasized in supervisory guidance. These themes are consistent with supervisory guidance, which emphasizes repayment capacity, leverage, liquidity, covenant discipline, refinancing constraints, and early recognition of weakening firms \citep{occ2001,fed2013,occ2024}. The PB Stress Score translates those recurring themes into a structured disclosure-based text measure. The PB Stress Score is therefore designed as a transparent textual operationalization of supervisory credit-risk mechanisms rather than as a generic tone measure.

Interpretability is also important in credit-risk applications because model outputs often need to be understood by analysts, validators, and governance functions. SHAP values, introduced by \citet{lundberg2017}, provide additive feature-level explanations of model predictions and have become widely used in credit-risk modeling. Recent credit-default studies also use SHAP to interpret ensemble models and identify the borrower characteristics that drive fitted predictions \citep{yang2025}. At the same time, \citet{lin2025}, for example, show that SHAP explanations can vary across model replications, especially for moderately important variables. This combination of applied use and stability caution supports the limited role assigned to SHAP in the present study: SHAP is used as an interpretability layer for fitted models, not as evidence of causality or as a tool for selecting the preferred specification.

\subsection{Contribution Relative to Prior Work}

The present study contributes to the literature in four ways. First, it defines the outcome from the 10-K filing date, aligning disclosure timing with the prediction window. Second, it evaluates text incrementally rather than in isolation by testing whether 10-K narratives improve rank-ordering value beyond a transparent accounting baseline. Third, it benchmarks the PB Stress Score against both named accounting models and the official Loughran--McDonald dictionary benchmark. Fourth, it operationalizes supervisory credit-risk mechanisms in a transparent and auditable text score rather than relying on a black-box model.

The contribution is therefore specific and transparent: structured distress signals in mandatory 10-K narrative sections can improve post-filing bankruptcy prediction beyond parsimonious accounting baselines.

\section{Data and Methodology}
\label{sec:method}

\subsection{Data Sources and Sample Selection}

This study combines three linked data sources to gather the financial variables, default date and status, and 10-K narratives. First, firm-year accounting variables and filing metadata are constructed from the U.S.\ Securities and Exchange Commission's (SEC) Electronic Data Gathering, Analysis, and Retrieval (EDGAR) filing system and structured SEC financial statement data \citep{sec2024}. These records provide the firm identifier, fiscal year, Standard Industrial Classification (SIC) industry classification, 10-K filing date, and financial statement variables used in the baseline bankruptcy-prediction models. Second, bankruptcy event information is obtained from the Florida--UCLA--LoPucki Bankruptcy Research Database (BRD), which provides externally identified corporate bankruptcy filing events and filing dates used to define the outcome variable \citep{brd}. Third, narrative disclosure data are extracted from matched 10-K filings, focusing on Item 1, Item 1A, and Item 7. The merged empirical sample is therefore organized at the firm-year filing level, with accounting data, filing metadata, bankruptcy outcomes, and extracted 10-K text aligned to the same 10-K observation.

The unit of observation is a firm-year 10-K filing, meaning one annual 10-K filing for one firm for one fiscal year. A firm can therefore contribute multiple pre-bankruptcy observations across different fiscal years, but observations dated after the firm's bankruptcy filing date are excluded from the prediction sample.

Because Form 10-K is filed by public reporting companies, the empirical universe is limited to public firms with SEC filing histories and usable structured financial data. Because bankruptcy filings are rare among public firms, the estimation sample is constructed from SIC major groups that contribute meaningful numbers of observable BRD bankruptcy events in the matched SEC--10-K universe. The resulting sample is therefore intended to support bankruptcy-risk analysis within a selected public-company universe rather than to represent all SEC registrants. The selected universe contains 15 SIC major groups and includes 10,778 unique SEC-covered firms across energy, manufacturing, transportation, communications, utilities, retail, financial and holding companies, business services, and health services.

The primary empirical sample is the final text-model sample: firm-year 10-K observations with an observable one-year bankruptcy outcome and usable extracted 10-K text. This final sample contains 40,475 firm-year observations from fiscal years 2010 through 2021, of which 128 are followed by a BRD-recorded bankruptcy filing within one year of the 10-K filing date.

\begin{table}[H]
\centering
\caption{Primary Sample Definition}
\label{tab:sample_definition}
\begin{tabularx}{\textwidth}{@{}p{0.28\textwidth}p{0.18\textwidth}p{0.12\textwidth}Y@{}}
\toprule
Measure & Unit & Count & Interpretation \\
\midrule
Selected SEC-covered firms & Firm & 10,778 & Unique number of firms in the 15 selected SIC major groups \\
One-year bankruptcy event observations & Firm-year 10-K observation & 128 & Bankruptcy filing occurs within 365 days of the 10-K filing date \\
Non-event observations & Firm-year 10-K observation & 40,347 & No BRD bankruptcy filing within the one-year prediction window \\
Final primary text-model sample & Firm-year 10-K observation & 40,475 & Observations with usable extracted text and observable one-year outcome \\
\bottomrule
\end{tabularx}
\end{table}

\subsection{Text Extraction and Bankruptcy Outcome Definition}

Consistent with the disclosure setting introduced in Section~\ref{sec:intro}, the text source is Form 10-K, the annual report filed by U.S.\ public companies with the SEC. The extraction process targets three narrative sections: Item 1, Item 1A, and Item 7. Item 1 describes the firm's business, operating structure, markets, and major business exposures. Item 1A describes risk factors, including financing, litigation, regulatory, customer, supplier, and operating risks. Item 7 contains Management's Discussion and Analysis, including management's discussion of performance, liquidity, capital resources, financing needs, and uncertainty. These sections are selected because they are the principal parts of the 10-K in which firms are expected to disclose material business conditions, risk factors, and liquidity or capital-resource pressures.

An observation is treated as having usable text if the combined extracted text from Item 1, Item 1A, and Item 7 contains at least 100 characters. This rule is intended to remove empty or failed extraction records without requiring every individual section to be present. Under this combined-section usable-text rule, the final primary text-model sample contains 40,475 observations, of which 128 are one-year bankruptcy event observations.

The primary dependent variable is an indicator for whether the firm files for bankruptcy within one year after the 10-K filing date. Specifically, the indicator equals one if the BRD bankruptcy filing date occurs from the 10-K filing date through 365 calendar days after the filing date, inclusive. It equals zero only if no BRD bankruptcy filing occurs during that one-year window. The one-year prediction window therefore refers to the 365-day interval beginning on the 10-K filing date, which is the point at which the filing becomes available to investors and creditors.

\subsection{Prediction Framework and Financial Baseline}

The objective is to test whether adding a 10-K text score improves the rank-ordering of firms by bankruptcy risk within the 365 calendar days after the 10-K filing date. All predictors are measured using information available at the 10-K filing date, and firm-year observations dated after a firm's bankruptcy filing date are excluded from the prediction sample.

The empirical design compares a financial-only logistic model with a text-augmented logistic model:
\begin{equation}
\text{logit}\!\left(P(Y = 1)\right) = \alpha + \beta_1 X_1 + \beta_2 X_2 + \cdots + \beta_k X_k
\end{equation}
\begin{equation}
\text{logit}\!\left(P(Y = 1)\right) = \alpha + \beta_1 X_1 + \beta_2 X_2 + \cdots + \beta_k X_k + \gamma \cdot \text{TextScore}
\end{equation}

The key empirical test is whether adding a pre-specified 10-K text score improves out-of-sample bankruptcy-risk ranking relative to the same accounting baseline estimated without the text score.

The primary financial baseline is a parsimonious five-factor accounting model based on firm size, leverage, liquidity, profitability, and operating cash flow. This baseline is used as the main accounting benchmark because it captures core borrower-risk dimensions emphasized in both the bankruptcy-prediction literature, supervisory credit analysis, and practitioner rating methodologies \citep{moodys2024,sp2021}. In this framework, firm size proxies for scale and resilience, leverage captures balance-sheet burden, liquidity captures short-term financial flexibility, profitability captures earnings capacity, and operating cash flow captures internally generated repayment capacity. The five-factor baseline therefore serves as a transparent and reproducible benchmark for one-year post-filing bankruptcy prediction.

\begin{table}[H]
\centering
\caption{Financial Baseline Variables}
\label{tab:financial_baseline}
\begin{tabularx}{\textwidth}{@{}p{0.2\textwidth}Y@{}}
\toprule
Variable & Definition \\
\midrule
Size & Natural log of total assets \\
Leverage & Total debt divided by total assets to represent balance-sheet leverage \\
Liquidity & Current assets divided by current liabilities \\
Profitability & Net income divided by total assets \\
Cash flow & Operating cash flow divided by total assets \\
\bottomrule
\end{tabularx}
\end{table}

These five variables provide a compact accounting-only benchmark for near-term bankruptcy prediction: larger firms, more liquid firms, more profitable firms, and firms with stronger operating cash flow are generally expected to face lower bankruptcy risk, while more highly leveraged firms are generally expected to face higher bankruptcy risk.

Two additional accounting benchmarks are also used: an Ohlson-style reduced score and a Zmijewski-style score, linked to the classic accounting-based models of \citet{ohlson1980} and \citet{zmijewski1984}. The Ohlson-style score combines variables related to firm size, leverage, working-capital structure, profitability, and negative equity. The Zmijewski-style score emphasizes profitability, leverage, and liquidity. In both benchmark implementations, a higher score indicates higher estimated bankruptcy risk. These measures are included as secondary accounting benchmarks. The exact benchmark-score constructions are reported in Appendix~\ref{app:benchmark_formulas}.

\subsection{Text Score Design}

The main analysis uses two text scores. The first is a Loughran--McDonald dictionary score, used as an established finance-text benchmark. The second is the PB Stress Score, the proposed primary pre-bankruptcy stress text score.

\begin{table}[H]
\centering
\caption{Main Text Scores}
\label{tab:text_scores}
\begin{tabularx}{\textwidth}{@{}p{0.2\textwidth}p{0.25\textwidth}Y@{}}
\toprule
Score & Role & Purpose \\
\midrule
LM text score & Benchmark text score & Measures general negative and uncertain tone using the finance-specific Loughran--McDonald dictionary \\
PB Stress Score & Proposed primary score & Measures pre-bankruptcy stress using a transparent dictionary-based five-pillar score \\
\bottomrule
\end{tabularx}
\end{table}

The paper's main comparison is therefore among three specifications: the financial-only baseline, the financial baseline plus the LM text benchmark, and the financial baseline plus the proposed PB Stress Score.

\subsubsection{Loughran--McDonald Dictionary Sentiment Score}

The LM text score is based on the official Loughran--McDonald financial dictionary \citep{lm_dict}. The Loughran--McDonald dictionary is a finance-specific word list developed from corporate filings because general-purpose sentiment dictionaries often misclassify common words in financial text \citep{lm2011}. It is widely used in accounting and finance research as a benchmark for measuring textual tone in firm disclosures \citep{lm2016}. The score uses the negative and uncertainty categories because these categories are well established in financial disclosure research and are better suited to finance text than generic sentiment dictionaries. Word counts are normalized by document length and expressed per 1,000 words. Because the prediction target is bankruptcy risk rather than general disclosure uncertainty, the benchmark is designed to place greater weight on negative language than on uncertainty language.

The LM-based negative-uncertainty composite benchmark is:
\begin{align}
\text{negative\_per\_1000} &= 1000 \cdot \frac{\text{LM\_negative\_word\_count}}{\text{total\_word\_count}} \\
\text{uncertainty\_per\_1000} &= 1000 \cdot \frac{\text{LM\_uncertainty\_word\_count}}{\text{total\_word\_count}} \\
\text{LM\_raw} &= 0.7 \cdot \text{negative\_per\_1000} + 0.3 \cdot \text{uncertainty\_per\_1000}
\end{align}

\begin{table}[H]
\centering
\caption{LM Example Word Table}
\label{tab:lm_example_words}
\begin{tabularx}{\textwidth}{@{}p{0.2\textwidth}Y@{}}
\toprule
LM Category & Examples \\
\midrule
Negative & adverse; decline; default; deficit; deteriorate; failure; impairment; loss; negative; terminate; unable \\
Uncertainty & approximate; approximately; assumed; contingent; estimate; fluctuate; possible; possibly; uncertainty; uncertain; unsure; variability \\
\bottomrule
\end{tabularx}
\end{table}

\subsubsection{Pre-Bankruptcy Stress Score}

The PB Stress Score is a pre-specified dictionary-based text score built from a frozen set of single-word and multi-word phrase patterns. The dictionary is organized around recurring pre-bankruptcy stress themes drawn from supervisory guidance issued by the OCC and the Federal Reserve, including repayment capacity, leverage, liquidity, refinancing dependence, debt maturities, covenant discipline, downside stress, and timely recognition of weakening firms \citep{occ2001,fed2013,occ2024}. These themes are operationalized into five disclosure pillars:
\begin{itemize}
    \item liquidity and funding stress
    \item debt, covenant, and refinancing stress
    \item operating deterioration
    \item restructuring and legal distress
    \item business fragility
\end{itemize}

Within each pillar, dictionary patterns are classified as severe, moderate, mild, or mitigating according to the intensity of the condition disclosed. Severe patterns indicate acute pre-bankruptcy stress, such as going-concern doubt, explicit inability to meet obligations, or default-like debt language. Moderate patterns indicate material but less terminal strain, such as covenant pressure, refinancing difficulty, recurring losses, impairment, or restructuring pressure. Mild patterns capture weaker background vulnerability signals, while mitigating patterns capture disclosures that offset stress, such as strong liquidity, covenant compliance, successful refinancing, or diversified customer and supplier exposure. Direct bankruptcy-status terms, including Chapter references and explicit statements that the firm filed for bankruptcy, are excluded from the PB dictionary.

Text is evaluated at the paragraph level. A dictionary pattern contributes one hit when it appears at least once in a paragraph; repeated occurrences of the same pattern within that paragraph count once, while different patterns in the same paragraph can each contribute one hit. To limit the influence of repetitive boilerplate, each pattern can contribute at most three paragraph-level hits per filing. Within a given pillar, \texttt{severe\_hits}, \texttt{moderate\_hits}, \texttt{mild\_hits}, and \texttt{mitigant\_hits} therefore represent the total number of category-specific pattern-paragraph matches across the filing.

Pillar points are then computed as:
\begin{equation}
\text{pillar\_points} = \max(4.0 \cdot \text{severe\_hits} + 2.0 \cdot \text{moderate\_hits} + 1.0 \cdot \text{mild\_hits} - 1.5 \cdot \text{mitigant\_hits}, 0)
\end{equation}

These coefficients impose an ordinal severity structure rather than a statistically estimated weighting scheme. Severe disclosures receive the largest weight, moderate disclosures receive a smaller weight, and mild disclosures receive the smallest weight so that the score distinguishes acute distress language from weaker background vulnerability signals. Mitigating disclosures receive a negative weight to offset part of the distress evidence, but the smaller absolute magnitude relative to severe disclosures prevents favorable language from fully canceling acute warning signals. The coefficients are therefore transparent calibration choices designed to preserve monotonicity and interpretability rather than to claim a unique optimal weighting.

Each pillar is then mapped to a bounded 0--100 component using:
\begin{equation}
\text{pillar\_component} = 100 \cdot \left(1 - \exp(-0.28 \cdot \text{pillar\_points})\right)
\end{equation}

The exponential transformation maps pillar points to a bounded 0--100 scale and imposes diminishing marginal effects from additional hits. The parameter 0.28 governs the rate of saturation and is chosen as a transparent calibration device so that early distress signals have meaningful impact while additional repetitive disclosures contribute at a decreasing rate.

The raw PB Stress Score is the equal-weight average of the five pillar components:
\begin{align}
\text{PB Stress Score Raw} &= 0.20 \cdot \text{liquidity\_funding\_component} \notag\\
&\quad + 0.20 \cdot \text{debt\_covenant\_refinancing\_component} \notag\\
&\quad + 0.20 \cdot \text{operating\_deterioration\_component} \notag\\
&\quad + 0.20 \cdot \text{restructuring\_legal\_component} \notag\\
&\quad + 0.20 \cdot \text{business\_fragility\_component}
\end{align}

Because each pillar component is bounded between 0 and 100, the raw PB Stress Score is also bounded between 0 and 100. For model estimation, text scores are normalized using percentile cutoffs estimated on the training sample and then held fixed for holdout and future observations, improving comparability across measures while limiting sensitivity to extreme values without altering the underlying signal extracted from each filing.

Table~\ref{tab:pb_phrase_categories} reports representative examples by pillar, while Appendix~\ref{app:pb_dictionary} reports the full frozen dictionary used in scoring.

\begin{table}[H]
\centering
\caption{Representative PB Stress Score Phrase Categories}
\label{tab:pb_phrase_categories}
\small
\begin{tabularx}{\textwidth}{@{}p{0.18\textwidth}YY@{}}
\toprule
Pillar & Representative Distress Signals & Examples of Mitigating Signals \\
\midrule
Liquidity and funding stress & substantial doubt; going concern; unable to meet obligations; liquidity shortfall; working capital deficit; negative operating cash flow; cash burn; need additional capital & sufficient liquidity; ample liquidity; available borrowing capacity; undrawn revolver; positive operating cash flow \\
Debt, covenant, and refinancing stress & event of default; notice of default; cross-default; debt acceleration; forbearance; covenant waiver; covenant breach; unable to refinance; debt maturity pressure; rating downgrade; distressed exchange & in compliance with covenants; no event of default; successfully refinanced; repaid debt; deleveraging \\
Operating deterioration & ceased operations; material decline in sales and cash flow; net loss; operating loss; negative EBITDA; impairment; write-down; restructuring charge; facility closure; layoffs; margin compression & net income; operating income; strong results; record revenue \\
Restructuring and legal distress & going-concern qualification; material weakness; debt restructuring; liability management; asset sales to raise liquidity; litigation contingency affecting liquidity & resolved litigation; remediated material weakness \\
Business fragility & customer loss; major customer termination; supplier disruption; customer concentration; supplier concentration; intense competition; price competition; cyclical demand; demand volatility & diversified customer base; diversified supplier base \\
\bottomrule
\end{tabularx}
\normalsize
\end{table}

\subsection{Descriptive Statistics}
\label{sec:descriptive_stats}

Table~\ref{tab:descriptive_stats} reports summary statistics for the final primary text-model sample used in the one-year analysis. This sample contains 40,475 firm-year 10-K observations with usable extracted text and observable one-year bankruptcy outcomes, including 128 bankruptcy events. The implied one-year event rate is therefore 0.32\%, which is consistent with the rare-event setting addressed in the estimation design.

The financial variables exhibit substantial heterogeneity, especially for liquidity, profitability, and operating cash flow, so Table~\ref{tab:descriptive_stats} reports medians and interquartile ranges in addition to means and standard deviations. The two text-based measures are reported in their model-input form after the preprocessing and normalization steps described above. The PB Stress Score has a higher mean and median than the LM benchmark in the estimation sample, although the predictive comparisons in Section~\ref{sec:results} remain the primary evidence on incremental value.

\begin{table}[H]
\centering
\caption{Descriptive Statistics for Main Model Variables}
\label{tab:descriptive_stats}
\small
\setlength{\tabcolsep}{5pt}
\begin{tabular}{@{}lrrrrrr@{}}
\toprule
Variable & N & Mean & Std. Dev. & P25 & Median & P75 \\
\midrule
\shortstack[l]{Size\\(log total assets, USD)} & 40,475 & 18.773 & 4.055 & 16.836 & 19.674 & 21.584 \\
\shortstack[l]{Leverage\\(total debt / total assets)} & 40,475 & 0.275 & 0.195 & 0.250 & 0.250 & 0.250 \\
\shortstack[l]{Liquidity\\(current assets / current liabilities)} & 40,475 & 2.659 & 3.840 & 1.057 & 1.626 & 2.445 \\
\shortstack[l]{Profitability\\(net income / total assets)} & 40,475 & -8.948 & 55.755 & -0.301 & 0.001 & 0.026 \\
\shortstack[l]{Operating cash flow\\(operating cash flow / total assets)} & 40,475 & -2.588 & 15.978 & -0.057 & 0.019 & 0.065 \\
\shortstack[l]{LM text score\\(negative--uncertainty benchmark)} & 40,475 & 51.033 & 20.337 & 38.036 & 52.274 & 64.670 \\
\shortstack[l]{PB Stress Score\\(distress-specific text score)} & 40,475 & 56.818 & 22.351 & 41.984 & 58.789 & 73.452 \\
\bottomrule
\end{tabular}

\vspace{0.3em}
\begin{minipage}{0.95\textwidth}
\footnotesize
\textit{Note.} Summary statistics are reported for the final one-year text-model sample with usable extracted 10-K text and observable one-year bankruptcy outcomes. The sample contains 40,475 observations, including 128 one-year bankruptcy events (0.32\%). The LM text score and PB Stress Score are reported in the model-input form used in estimation after preprocessing and normalization.
\end{minipage}
\normalsize
\end{table}

\subsection{Model Evaluation and Interpretation}

Models are estimated using L2-penalized logistic regression with inverse-frequency class weighting because bankruptcy is a rare event. Recent work on low-prevalence financial classification problems also emphasizes that training objectives and evaluation choices can materially affect performance under extreme imbalance \citep{sun2025}. Performance is evaluated primarily using rank-ordering metrics because the objective is to determine whether 10-K narrative information improves the ability to distinguish firm-year observations that later file for bankruptcy from those that do not. The primary evaluation metrics are AUC, Somers' D, and the Kolmogorov--Smirnov statistic (KS). PR-AUC and top-decile capture are reported as secondary metrics \citep{davis2006}. Models are estimated on a fixed roughly 80/20 training--holdout split. The exact number of usable observations varies slightly across specifications because some filings do not yield usable text-based measures.

AUC measures the probability that a randomly selected bankruptcy observation receives a higher predicted risk score than a randomly selected non-bankruptcy observation \citep{fawcett2006}. Somers' D is a linear transformation of AUC, defined as $\text{Somers' D} = 2 \times \text{AUC} - 1$. The KS statistic measures the maximum separation between the cumulative predicted-score distributions of bankruptcy and non-bankruptcy observations. Higher values of AUC, Somers' D, and KS indicate stronger rank-ordering performance.

Incremental value is assessed by comparing each text-augmented model with its corresponding financial-only baseline on the same eligible observations. The main incremental metrics are:
\begin{align}
\Delta \text{AUC} &= \text{AUC}(\text{financial baseline + text score}) - \text{AUC}(\text{financial baseline}) \\
\Delta \text{Somers' D} &= \text{Somers' D}(\text{financial baseline + text score}) - \text{Somers' D}(\text{financial baseline}) \\
\Delta \text{KS} &= \text{KS}(\text{financial baseline + text score}) - \text{KS}(\text{financial baseline})
\end{align}

Paired bootstrap confidence intervals are used for the incremental metrics \citep{efron1993}. Each specification is fit once on a fixed roughly 80/20 training--holdout split, and the bootstrap is applied to the eligible holdout firm-year observations only. In each of 2,000 replicates, holdout observations are resampled with replacement, and the metric difference between the financial-only model and the corresponding text-augmented model is recomputed on the same resampled observations. Because resampling is performed at the firm-year level and the model is not refit within each replicate, the intervals quantify holdout-sample uncertainty in incremental performance but do not fully capture within-firm dependence.

To supplement the predictive-performance tests, the study also reports SHAP-style feature contributions for the fitted logistic models \citep{lundberg2017}. The contribution of feature $j$ for observation $i$ is computed as:
\begin{equation}
\text{SHAP}_{ij} = \beta_j \times z_{ij}
\end{equation}
where $\beta_j$ is the fitted coefficient and $z_{ij}$ is the standardized value of feature $j$ for observation $i$.

Feature importance is summarized by the mean absolute contribution of each feature across observations. The SHAP results are used to describe how the fitted model uses the text score relative to the financial variables, while predictive-performance tests remain the primary evidence on incremental value.

\section{Results}
\label{sec:results}

\subsection{Primary One-Year Bankruptcy Prediction Results}

Table~\ref{tab:primary_results} reports the main results for one-year post-filing bankruptcy prediction. The primary comparison is between the five-variable financial baseline, the same baseline augmented with the Loughran--McDonald benchmark, and the same baseline augmented with the PB Stress Score. Performance is evaluated primarily using AUC, Somers' D, and KS, with PR-AUC and top-decile capture reported as secondary metrics.

The five-variable financial baseline achieves an AUC of 0.8323, a Somers' D of 0.6646, and a KS statistic of 0.5246. Adding the Loughran--McDonald benchmark improves AUC to 0.8602 and KS to 0.5673, indicating that general negative and uncertainty language contains some incremental information about subsequent bankruptcy risk, although its PR-AUC is slightly lower than that of the baseline model. Adding the PB Stress Score raises AUC further to 0.9019, Somers' D to 0.8037, and KS to 0.6511. The PB Stress Score also increases PR-AUC from 0.0534 to 0.1021 and raises top-decile capture from 44.12\% to 64.71\%. These results indicate that structured pre-bankruptcy stress disclosures provide materially stronger rank-ordering value than the broad dictionary benchmark.

\begin{table}[H]
\centering
\caption{Primary One-Year Holdout Results}
\label{tab:primary_results}
\small
\begin{tabular}{@{}lccccc@{}}
\toprule
Model & AUC & Somers' D & KS & PR-AUC & Top-Decile Capture \\
\midrule
Five-variable financial baseline & 0.8323 & 0.6646 & 0.5246 & 0.0534 & 44.12\% \\
Financial baseline + LM text score & 0.8602 & 0.7204 & 0.5673 & 0.0507 & 50.00\% \\
Financial baseline + PB Stress Score & 0.9019 & 0.8037 & 0.6511 & 0.1021 & 64.71\% \\
\bottomrule
\end{tabular}

\vspace{0.3em}
\begin{minipage}{0.95\textwidth}
\footnotesize
\textit{Note.} The financial-only baseline uses 7,658 holdout observations. The two text-augmented specifications use 7,647 holdout observations because some filings do not yield usable text-based measures. All three specifications contain 34 holdout bankruptcy observations.
\end{minipage}
\normalsize
\end{table}

\subsection{Incremental Value of the PB Stress Score}

The incremental value of the PB Stress Score is evaluated by comparing the text-augmented model with the corresponding financial-only baseline on the same holdout observations. Table~\ref{tab:bootstrap_results} reports paired bootstrap results for the primary one-year specification.

In the main holdout sample, the PB Stress Score increases AUC by 0.0696, with a paired bootstrap 95\% confidence interval of [0.0349, 0.1114]. The corresponding increase in Somers' D is 0.1393, with a confidence interval of [0.0697, 0.2227], and the increase in KS is 0.1266, with a confidence interval of [0.0442, 0.2189]. The bootstrap probability that the incremental effect is positive is 1.0000 for AUC and Somers' D and 0.9995 for KS. The evidence therefore supports the conclusion that the PB Stress Score provides statistically meaningful incremental rank-ordering value beyond the five-variable financial baseline.

\begin{table}[H]
\centering
\caption{Paired Bootstrap Tests of Incremental PB Stress Score Value}
\label{tab:bootstrap_results}
\small
\begin{tabular}{@{}lccc@{}}
\toprule
Metric & Observed Delta & 95\% Bootstrap CI & Prob(Delta $>$ 0) \\
\midrule
$\Delta$AUC & 0.0696 & [0.0349, 0.1114] & 1.0000 \\
$\Delta$Somers' D & 0.1393 & [0.0697, 0.2227] & 1.0000 \\
$\Delta$KS & 0.1266 & [0.0442, 0.2189] & 0.9995 \\
\bottomrule
\end{tabular}
\normalsize
\end{table}

\subsection{Alternative Accounting Benchmarks}

To assess whether the main result depends on the five-variable financial baseline, the PB Stress Score is also evaluated against Ohlson-style and Zmijewski-style reduced accounting benchmarks. These are secondary benchmark implementations linked to the classic bankruptcy-prediction literature and are reported as robustness evidence rather than as the paper's primary accounting comparison.

Table~\ref{tab:alternative_benchmarks} shows that the PB Stress Score remains strongly informative under both alternative accounting specifications. In the one-year holdout sample, the Ohlson-style benchmark alone produces an AUC of 0.4541, while the Zmijewski-style benchmark produces an AUC of 0.5894. Adding the PB Stress Score raises these AUC values to 0.8332 and 0.8024, respectively. The corresponding paired-bootstrap AUC improvements are 0.3789 for the Ohlson-style benchmark and 0.2133 for the Zmijewski-style benchmark, with both confidence intervals well above zero.

These results show that the PB Stress Score remains informative under additional accounting benchmark specifications, although the primary inference of the paper continues to rest on the stronger five-variable financial baseline.

\begin{table}[H]
\centering
\caption{Alternative Accounting Benchmarks and PB Stress Score Augmentation}
\label{tab:alternative_benchmarks}
\small
\setlength{\tabcolsep}{4pt}
\begin{tabular}{@{}lcccccc@{}}
\toprule
\shortstack{Baseline\\Specification} &
\shortstack{Baseline\\AUC} &
\shortstack{Baseline + PB\\AUC} &
$\Delta$AUC &
\shortstack{95\% Bootstrap\\CI} &
\shortstack{Baseline\\Top-Decile} &
\shortstack{Baseline + PB\\Top-Decile} \\
\midrule
\shortstack[l]{Ohlson-style\\reduced benchmark} & 0.4541 & 0.8332 & 0.3789 & [0.2789, 0.4740] & 0.00\% & 52.94\% \\
\shortstack[l]{Zmijewski-style\\reduced benchmark} & 0.5894 & 0.8024 & 0.2133 & [0.1520, 0.2713] & 2.94\% & 41.18\% \\
\bottomrule
\end{tabular}
\normalsize
\end{table}

\subsection{Robustness Tests}

\subsubsection{Robustness Tests on Alternative Bankruptcy Outcome Definitions}

The main result remains positive under alternative bankruptcy-outcome definitions. The first robustness test extends the outcome window from one year to two years after the 10-K filing date. The second excludes bankruptcies that occur within 30 days after the filing date, providing a stricter test of whether the text score is informative beyond filings immediately preceding bankruptcy.

As reported in Table~\ref{tab:outcome_robustness}, the five-variable financial baseline achieves an AUC of 0.8416 under the two-year outcome, while the corresponding PB Stress Score model achieves an AUC of 0.8917. The paired-bootstrap AUC improvement is 0.0501, with a 95\% confidence interval of [0.0178, 0.0828]. When bankruptcies within the first 30 days are excluded, the baseline AUC is 0.8281 and the PB Stress Score model achieves 0.8978, corresponding to an AUC improvement of 0.0697 with a 95\% confidence interval of [0.0331, 0.1131]. The corresponding Somers' D and KS differences are also positive in both robustness settings, although the two-year KS interval is weaker than the one-year results. Overall, the robustness tests support the interpretation that the PB Stress Score captures pre-bankruptcy stress signals that remain informative under alternative timing definitions.

\begin{table}[H]
\centering
\caption{Robustness to Alternative Bankruptcy Outcome Definitions}
\label{tab:outcome_robustness}
\small
\setlength{\tabcolsep}{3pt}
\begin{tabular}{@{}lccccccc@{}}
\toprule
\shortstack{Outcome\\Definition} &
\shortstack{Base\\AUC} &
\shortstack{Base +\\LM} &
\shortstack{Base +\\PB} &
\shortstack{$\Delta$AUC\\(PB--Base)} &
\shortstack{95\%\\CI} &
\shortstack{Base\\Top-Decile} &
\shortstack{PB\\Top-Decile} \\
\midrule
\shortstack[l]{Bankruptcy within 2 years\\after filing} & 0.8416 & 0.8659 & 0.8917 & 0.0501 & [0.0178, 0.0828] & 46.51\% & 60.47\% \\
\shortstack[l]{Bankruptcy within 1 year,\\excluding first 30 days} & 0.8281 & 0.8554 & 0.8978 & 0.0697 & [0.0331, 0.1131] & 43.75\% & 62.50\% \\
\bottomrule
\end{tabular}

\vspace{0.3em}
\begin{minipage}{0.95\textwidth}
\footnotesize
\textit{Note.} ``Base'' denotes the five-variable financial baseline.
\end{minipage}
\normalsize
\end{table}

\subsubsection{Robustness Tests on Out-of-Time Validation}

The incremental value of the PB Stress Score also remains visible in out-of-time validation. These tests re-estimate the models on earlier data and evaluate them on later yearly test sets, thereby assessing whether the main result persists outside the single primary holdout split.

Table~\ref{tab:oot_results} reports weighted out-of-time performance across the yearly test windows. For readability, the table is presented in two panels. Panel A reports weighted AUC and weighted Somers' D. Panel B reports weighted KS and top-decile capture. In the primary one-year setting, the five-variable financial baseline achieves a weighted AUC of 0.8419, compared with 0.8884 for the PB Stress Score model. Weighted Somers' D increases from 0.6838 to 0.7767, and weighted KS increases from 0.6125 to 0.7336. Weighted top-decile capture rises from 54.52\% to 69.92\%. Similar positive patterns appear under the two-year outcome and the one-year outcome excluding the first 30 days. The out-of-time results therefore support the interpretation that the PB Stress Score captures a stable disclosure signal rather than a result confined to the primary holdout split.

\begin{table}[H]
\centering
\caption{Out-of-Time Validation Results}
\label{tab:oot_results}
\small
\setlength{\tabcolsep}{4pt}

\begin{tabular}{@{}lcccc@{}}
\toprule
\multicolumn{5}{@{}l@{}}{\textit{Panel A. Weighted AUC and Weighted Somers' D}} \\
\midrule
\shortstack{Outcome\\Definition} &
\shortstack{Baseline\\Weighted AUC} &
\shortstack{PB\\Weighted AUC} &
\shortstack{Baseline\\Weighted Somers' D} &
\shortstack{PB\\Weighted Somers' D} \\
\midrule
\shortstack[l]{Bankruptcy within 1 year\\after filing} & 0.8419 & 0.8884 & 0.6838 & 0.7767 \\
\shortstack[l]{Bankruptcy within 2 years\\after filing} & 0.8269 & 0.8549 & 0.6539 & 0.7098 \\
\shortstack[l]{Bankruptcy within 1 year,\\excluding first 30 days} & 0.8303 & 0.8753 & 0.6606 & 0.7506 \\
\bottomrule
\end{tabular}

\vspace{0.8em}

\begin{tabular}{@{}lcccc@{}}
\toprule
\multicolumn{5}{@{}l@{}}{\textit{Panel B. Weighted KS and Top-Decile Capture}} \\
\midrule
\shortstack{Outcome\\Definition} &
\shortstack{Baseline\\Weighted KS} &
\shortstack{PB\\Weighted KS} &
\shortstack{Baseline\\Top-Decile} &
\shortstack{PB\\Top-Decile} \\
\midrule
\shortstack[l]{Bankruptcy within 1 year\\after filing} & 0.6125 & 0.7336 & 54.52\% & 69.92\% \\
\shortstack[l]{Bankruptcy within 2 years\\after filing} & 0.5409 & 0.6053 & 47.56\% & 57.93\% \\
\shortstack[l]{Bankruptcy within 1 year,\\excluding first 30 days} & 0.6166 & 0.7160 & 51.29\% & 64.35\% \\
\bottomrule
\end{tabular}
\normalsize
\end{table}

\subsection{SHAP-Based Interpretability}

Consistent with the evaluation framework defined earlier, SHAP-style log-odds contributions are reported for the primary one-year holdout model that augments the financial baseline with the PB Stress Score. These values are descriptive and summarize how the fitted model uses each feature; they are not interpreted as causal effects.

In the Financial baseline + PB Stress Score model, the PB Stress Score is the largest contributor by mean absolute SHAP value (0.5518), ranking first among all six features. It accounts for 38.79\% of total mean absolute SHAP contribution, with a bootstrap 95\% interval of [38.26\%, 39.32\%]. Its contribution is 2.88 times the median financial feature and 1.50 times the largest financial feature. These results indicate that the fitted model relies materially on the PB Stress Score when ranking observations by bankruptcy risk, while the predictive-performance tests in the earlier sections remain the primary evidence on incremental value.

\begin{table}[H]
\centering
\caption{SHAP Feature Importance for Financial Baseline + PB Stress Score (Primary One-Year Holdout)}
\label{tab:shap_importance}
\small
\begin{tabular}{@{}clcc@{}}
\toprule
Rank & Feature & Mean Abs SHAP (logit) & Share (\%) \\
\midrule
1 & PB Stress Score & 0.5518 & 38.79 \\
2 & Size (\texttt{size\_log\_assets}) & 0.3689 & 25.93 \\
3 & Leverage (\texttt{leverage\_debt\_to\_assets}) & 0.2820 & 19.82 \\
4 & Liquidity (\texttt{liquidity\_current\_ratio}) & 0.1918 & 13.48 \\
5 & Profitability (\texttt{profitability\_roa}) & 0.0143 & 1.01 \\
6 & Operating cash flow (\texttt{cashflow\_ocf\_to\_assets}) & 0.0137 & 0.96 \\
\bottomrule
\end{tabular}
\normalsize
\end{table}

\subsection{Summary of Main Findings}

The evidence in this section supports three main conclusions. First, 10-K narrative disclosures contain meaningful incremental information for predicting bankruptcy filings within one year after the filing date. Second, the PB Stress Score provides stronger rank-ordering value than the broad Loughran--McDonald benchmark in the primary holdout sample. Third, the result remains visible under alternative accounting benchmarks, alternative bankruptcy-outcome definitions, and out-of-time validation. Taken together, the results indicate that structured pre-bankruptcy stress language in 10-K filings can complement conventional accounting variables in bankruptcy-risk monitoring.

\section{Discussion and Conclusion}
\label{sec:conclusions}

This study asks whether narrative disclosures in annual 10-K filings provide incremental information for bankruptcy-risk ranking beyond the accounting variables disclosed in the same filing. The evidence indicates that they do. Although the five-variable financial baseline remains a strong benchmark, adding the PB Stress Score materially improves one-year post-filing bankruptcy prediction. This improvement is not confined to the primary holdout sample: the result remains positive under alternative accounting benchmarks, alternative bankruptcy-outcome definitions, a near-event exclusion test, and out-of-time validation. Taken together, the results suggest that structured 10-K narratives contain distress information that is not fully summarized by standard financial ratios.

The interpretation is economic rather than purely tonal. The PB Stress Score is designed to capture disclosure patterns tied to specific pre-bankruptcy mechanisms, including liquidity and funding pressure, debt covenant strain, refinancing difficulty, operating deterioration, restructuring and legal stress, and business fragility. This design distinguishes the score from broad dictionary-based tone measures that capture general negative or uncertain language. The results therefore suggest that bankruptcy prediction benefits not simply from measuring whether firms sound negative, but from identifying whether firms disclose concrete stress channels that are economically related to default risk.

The findings also help clarify the role of narrative disclosure within bankruptcy prediction. In this setting, both accounting variables and 10-K narratives become available at the same filing date, so the text score does not gain an informational advantage by using later data. Instead, its incremental value appears to come from measuring dimensions of firm stress that are not well summarized by the accounting baseline alone. Accounting ratios capture reported financial outcomes, while narrative disclosure can reveal the pathways through which financial stress is developing. A distress-specific text score built around those pathways can therefore improve interpretable bankruptcy-risk ranking beyond accounting variables alone.

The practical implication is additive use, not replacement. Accounting models remain essential because financial ratios provide standardized and interpretable measures of firm condition. The PB Stress Score is better viewed as a complementary surveillance signal for watchlist triage, early-warning review, and risk-based prioritization. It should not be interpreted as a standalone bankruptcy decision rule or as a fully calibrated probability-of-default model. Consistent with this interpretation, the SHAP-based analysis shows that the PB Stress Score is an important feature in the fitted text-augmented model, but that evidence remains descriptive rather than causal.

Several limitations remain. First, the outcome is bankruptcy filing rather than the broader default spectrum, so the results should not be interpreted as covering all forms of credit deterioration. Second, bankruptcy events are rare, which limits precision despite the large overall sample. Third, the text score can use only what firms choose or are required to disclose, so undisclosed risks remain outside the model. Fourth, the bootstrap is implemented at the firm-year rather than firm level, which means the reported intervals do not fully capture within-firm dependence. Finally, the Ohlson-style and Zmijewski-style benchmarks are reduced implementations included for robustness rather than full replications of the original published specifications.

Future work could extend the framework to broader default outcomes, additional disclosure sources, industry-specific dictionaries, and comparisons with more flexible NLP/LLM, multi-LLM, and retrieval-augmented systems for long financial reports, as well as LLM-based extraction frameworks for other disclosure types, benchmark-driven evaluations of financial reasoning performance, and potentially multimodal sentiment architectures when disclosure text is combined with other information channels, under strict anti-leakage controls \citep{cheng2026a,liu2026,kang2026,zhang2026,fu2026}. More broadly, the results support a simple conclusion: distress-specific 10-K narrative disclosures can complement conventional accounting ratios in bankruptcy-risk monitoring, and transparent text measures organized around recognizable credit-risk mechanisms can add meaningful predictive value without sacrificing interpretability.

\medskip
\noindent\textbf{Funding.} This research received no external funding.

\medskip
\noindent\textbf{Competing interests.} The authors declare no competing interests.

\clearpage
\appendix

\section{Benchmark Score Formulas and Empirical Implementations}
\label{app:benchmark_formulas}

This appendix reports the benchmark accounting-score formulas used for comparison in the paper. To avoid confusion between published bankruptcy-prediction models and the empirical benchmark implementations used here, the appendix distinguishes between the original published formulas and the reduced benchmark specifications constructed from the variables available in the final merged filing panel. The purpose of these benchmark scores is comparative rather than archival replication. The paper's primary accounting benchmark remains the five-variable financial baseline described in Section~3.3.

\subsection{Original Ohlson O-Score Formula}

For reference, the original Ohlson (1980) model is a nine-variable logit specification. The published score is:
\begin{equation}
\begin{aligned}
O =\;& -1.32 - 0.407 \cdot \mathrm{SIZE} + 6.03 \cdot \mathrm{TLTA} - 1.43 \cdot \mathrm{WCTA} \\
& + 0.0757 \cdot \mathrm{CLCA} - 1.72 \cdot \mathrm{OENEG} - 2.37 \cdot \mathrm{NITA} \\
& - 1.83 \cdot \mathrm{FUTL} + 0.285 \cdot \mathrm{INTWO} - 0.521 \cdot \mathrm{CHIN}
\end{aligned}
\end{equation}

where:
\begin{description}
  \item[SIZE] $\log(\text{TA} / \text{GNP price-level index})$
  \item[TLTA] total liabilities / total assets
  \item[WCTA] working capital / total assets
  \item[CLCA] current liabilities / current assets
  \item[OENEG] 1 if total liabilities exceed total assets, and 0 otherwise
  \item[NITA] net income / total assets
  \item[FUTL] funds from operations / total liabilities
  \item[INTWO] 1 if net income was negative for the last two years, and 0 otherwise
  \item[CHIN] $(NI_t - NI_{t-1}) / (|NI_t| + |NI_{t-1}|)$
\end{description}

The implied bankruptcy probability is:
\begin{equation}
P(\text{bankruptcy}) = \frac{1}{1 + \exp(-O)}
\end{equation}

This equation is reported for reference to the original published model. The empirical benchmark implementation used in this paper is described separately in Appendix~A.3.

\subsection{Original Zmijewski Formula}

For reference, the original Zmijewski (1984) model is a probit specification using three accounting variables. The published score is:
\begin{equation}
\text{ZMJ} = -4.3 - 4.5 \cdot \text{ROA} + 5.7 \cdot \text{TLTA} - 0.004 \cdot \text{CACL}
\end{equation}

where:
\begin{description}
  \item[ROA] net income / total assets
  \item[TLTA] total liabilities / total assets
  \item[CACL] current assets / current liabilities
\end{description}

The implied bankruptcy probability is obtained from the standard normal cumulative distribution function:
\begin{equation}
P(\text{bankruptcy}) = \Phi(\text{ZMJ})
\end{equation}

This equation is reported for reference to the original published model. The empirical benchmark implementation used in this paper is described separately in Appendix~A.3.

\subsection{Empirical Benchmark Implementations Used in This Paper}

The paper uses Ohlson-style and Zmijewski-style benchmark scores rather than claiming full exact replications of the original published models. Some original variables, transformations, and data definitions are not directly available in the final merged SEC filing panel used in this study. The benchmark implementations below are therefore designed to preserve the core accounting dimensions of the original models while maintaining consistency with the available data.

The empirical Ohlson-style benchmark used in this paper is:
\begin{equation}
\text{Ohlson\_style\_reduced\_score} = \frac{100}{1 + \exp(-X_{\text{ohlson}})}
\end{equation}

with
\begin{equation}
X_{\text{ohlson}} = -1.32 - 0.407 \cdot \text{SIZE} + 6.03 \cdot \text{TLTA} - 1.43 \cdot \text{WCTA} + 0.0757 \cdot \text{CLCA} - 2.37 \cdot \text{NITA} + 1.72 \cdot \text{OENEG}
\end{equation}

where:
\begin{description}
  \item[SIZE] natural log of total assets
  \item[TLTA] total liabilities / total assets
  \item[WCTA] working capital / total assets
  \item[CLCA] current liabilities / current assets
  \item[NITA] net income / total assets
  \item[OENEG] 1 if shareholders' equity is negative, and 0 otherwise
\end{description}

The empirical Zmijewski-style benchmark used in this paper is:
\begin{equation}
\text{Zmijewski\_style\_score} = 100 \cdot \Phi(X_{\text{zmijewski}})
\end{equation}

with
\begin{equation}
X_{\text{zmijewski}} = -4.3 - 4.5 \cdot \text{NITA} + 5.7 \cdot \text{TLTA} - 0.004 \cdot \text{CACL}
\end{equation}

where:
\begin{description}
  \item[NITA] net income / total assets
  \item[TLTA] total liabilities / total assets
  \item[CACL] current assets / current liabilities
  \item[$\Phi$] standard normal cumulative distribution function
\end{description}

In both empirical benchmark implementations, a higher score indicates higher estimated bankruptcy risk. These measures are included as literature-linked accounting comparators. The primary accounting specification in the paper remains the five-variable financial baseline reported in Section~3.3.

\begin{table}[H]
\centering
\caption{Benchmark Formula Summary}
\label{tab:benchmark_formula_summary}
\begin{tabularx}{\textwidth}{@{}p{0.2\textwidth}p{0.22\textwidth}p{0.2\textwidth}Y@{}}
\toprule
Benchmark & Type & Formula Basis & Notes \\
\midrule
Ohlson O-Score & Original published model & Nine-variable logit & Reported for reference in Appendix~A.1 \\
Ohlson-style reduced score & Empirical implementation used in this paper & Reduced logistic benchmark using available variables & Reported in Appendix~A.3 \\
Zmijewski model & Original published model & Three-variable probit & Reported for reference in Appendix~A.2 \\
Zmijewski-style score & Empirical implementation used in this paper & Probit-style benchmark using available variables & Reported in Appendix~A.3 \\
\bottomrule
\end{tabularx}
\end{table}

\section{PB Stress Score Dictionary}
\label{app:pb_dictionary}

\begin{longtable}{@{}p{0.18\textwidth}p{0.76\textwidth}@{}}
\caption{Liquidity and Funding Stress Dictionary}\label{tab:pb_liquidity_dict}\\
\toprule
Bucket & Canonical phrase(s) \\
\midrule
\endfirsthead
\toprule
Bucket & Canonical phrase(s) \\
\midrule
\endhead
\bottomrule
\endfoot
Severe & substantial doubt \\
Severe & ability to continue as a going concern \\
Severe & unable or cannot meet or satisfy obligations, debt service, or commitments \\
Moderate & liquidity constraint; liquidity shortfall; lack of liquidity \\
Moderate & working capital deficit; negative working capital \\
Moderate & negative operating cash flow; net cash used in operating activities \\
Moderate & cash burn; burn rate \\
Moderate & need additional capital; raise additional capital \\
Mild & liquidity \\
Mild & fund operations; finance operations \\
Mitigating & sufficient liquidity; ample liquidity; strong liquidity; adequate liquidity \\
Mitigating & net cash provided by operating activities; positive operating cash flow \\
Mitigating & available borrowing capacity; undrawn credit facility; undrawn revolver \\
\end{longtable}

\begin{longtable}{@{}p{0.18\textwidth}p{0.76\textwidth}@{}}
\caption{Debt, Covenant, and Refinancing Stress Dictionary}\label{tab:pb_debt_dict}\\
\toprule
Bucket & Canonical phrase(s) \\
\midrule
\endfirsthead
\toprule
Bucket & Canonical phrase(s) \\
\midrule
\endhead
\bottomrule
\endfoot
Severe & event of default; notice of default; default under a credit, debt, or loan agreement \\
Severe & cross-default; acceleration of debt \\
Severe & forbearance agreement; covenant waiver; waiver from lenders \\
Moderate & covenant breach; breach of covenant; violation of covenant \\
Moderate & unable or cannot refinance \\
Moderate & debt maturities; notes due in a stated year \\
Moderate & credit rating downgrade; downgrade of credit rating \\
Moderate & debt restructuring; distressed exchange \\
Mild & covenant; covenants \\
Mild & credit facility; revolver; term loan; senior notes \\
Mitigating & in compliance with covenants \\
Mitigating & no event of default; not in default \\
Mitigating & successfully refinanced; completed refinancing \\
Mitigating & repaid debt; paid down debt; deleveraging \\
\end{longtable}

\begin{longtable}{@{}p{0.18\textwidth}p{0.76\textwidth}@{}}
\caption{Operating Deterioration Dictionary}\label{tab:pb_operating_dict}\\
\toprule
Bucket & Canonical phrase(s) \\
\midrule
\endfirsthead
\toprule
Bucket & Canonical phrase(s) \\
\midrule
\endhead
\bottomrule
\endfoot
Severe & going out of business; ceased operations \\
Severe & material decline in sales or revenue and cash flow \\
Moderate & net loss; operating loss \\
Moderate & negative EBITDA \\
Moderate & impairment charge; asset impairment; goodwill impairment; write-down \\
Moderate & restructuring charge; restructuring plan \\
Moderate & store closures; facility closure; plant closure \\
Moderate & layoffs; reduction in force; workforce reduction \\
Moderate & margin compression; gross margin declined; gross margin decreased \\
Mild & decline in sales; decline in revenue; decline in demand; decline in orders \\
Mild & adverse trend; deterioration \\
Mitigating & net income; operating income; profitability \\
Mitigating & record sales; record revenue; record earnings; strong results; strong performance \\
\end{longtable}

\begin{longtable}{@{}p{0.18\textwidth}p{0.76\textwidth}@{}}
\caption{Restructuring and Legal Distress Dictionary}\label{tab:pb_restructuring_dict}\\
\toprule
Bucket & Canonical phrase(s) \\
\midrule
\endfirsthead
\toprule
Bucket & Canonical phrase(s) \\
\midrule
\endhead
\bottomrule
\endfoot
Severe & going-concern qualification \\
Severe & material weakness in internal control \\
Moderate & debt restructuring; liability management \\
Moderate & asset sale to raise liquidity; sale of assets to fund operations \\
Moderate & litigation reserve; legal contingencies that could impact liquidity \\
Mild & reorganization; strategic alternatives; cost reduction program \\
Mitigating & resolved litigation; no material adverse impact from litigation \\
Mitigating & remediated material weakness \\
\end{longtable}

\begin{longtable}{@{}p{0.18\textwidth}p{0.76\textwidth}@{}}
\caption{Business Fragility Dictionary}\label{tab:pb_business_dict}\\
\toprule
Bucket & Canonical phrase(s) \\
\midrule
\endfirsthead
\toprule
Bucket & Canonical phrase(s) \\
\midrule
\endhead
\bottomrule
\endfoot
Severe & significant customer loss; termination of major customer relationship \\
Severe & supplier disruption that materially impacted operations \\
Moderate & customer concentration; single customer represents a material portion \\
Moderate & supplier concentration; single source supplier \\
Moderate & highly competitive; intense competition; price competition \\
Moderate & cyclical demand; demand volatility \\
Mild & exposure to commodity price volatility; foreign exchange volatility \\
Mitigating & diversified customer base; no single customer accounts for a material portion \\
Mitigating & diversified supplier base \\
\end{longtable}

\end{document}